# New evaluation of neutron lifetime from UCN storage experiments and beam experiments

A.P. Serebrov*, A.K. Fomin

*Petersburg Nuclear Physics Institute, Russian Academy of Sciences, RU-188300 Gatchina, Leningrad District, Russia*

**Abstract**

The analysis of experiments on measuring neutron lifetime has been made. The latest most accurate result of measuring neutron lifetime [Phys. Lett. B 605, 72 (2005)] 878.5 ± 0.8 s differs from the world average value [Phys. Lett. B 667, 1 (2008)] 885.7 ± 0.8 s by 6.5 standard deviations. In view of this both the analysis and the Monte Carlo simulation of experiments [Phys. Lett. B 483, 15 (2000)] and [Phys. Rev. Lett. 63, 593 (1989)] have been performed. Systematic errors about −6 s have been found in both experiments. The table of results of neutron lifetime measurements is given after corrections and additions have been made. A new world average value of neutron lifetime makes up 880.0 ± 0.9 s. Here is also presented a separate analysis of experiments on measuring neutron lifetime with UCN and experiments on the beams. The average neutron lifetime for experiments with UCN is equal to 879.3(0.6) s, while for experiments on the beams it is equal to 889.1(2.9) s. The present difference of average values for both groups is (3.3σ) and needs consideration. The contribution of beam experiments into the world average value is not high, therefore it does not influence the above analysis. However, it is an independent problem to be solved. It seems desirable that the precision of beam experiments should be enhanced.

A new most accurate result of measuring neutron lifetime [1] 878.5 ± 0.8 c differs from the world average value [2] 885.7 ± 0.8 c by 6.5 standard deviations. The experiment [1] used a gravitational trap of ultracold neutrons (UCN) with covering made of the low-temperature fluorinated oil that has a few advantages over the previous experiments:

1) The coefficient of UCN losses for such a covering makes up only $2 \cdot 10^{-6}$, which leads to the probability of neutron losses in collisions with walls being only 1% of β-decay probability. In such a way one could observe almost direct process of neutron decay in the trap.

2) Extrapolation from the best storage time to neutron lifetime is 5 s. An experimental error is equal to ± $0.7_{stat}$ ± $0.3_{syst}$ s, i.e. relative accuracy of determining losses at wall collisions is 10%. In such conditions it is practically impossible to get a systematic error 7 s.

3) In using a low-temperature fluorinated oil the quasi-elastic scattering of UCN is completely suppressed. The upper constraint on possible correction concerned with this process is 0.03 s.

4) There is no effect of stationary trajectories of UCN.

5) Owing to high surface properties the surface of traps without covering is less than $10^{-6}$. It ensures identity of factor of losses for different traps.

6) The course of the experiment has demonstrated stability and reproducibility of trap coverings.

The situation is rather dramatic in estimating the world average value of neutron lifetime. On the one hand, a new value of neutron lifetime from the paper [1] cannot be included in the world average value because of great difference in results. On the other hand, new data with high measurement precision makes doubtful the world average value for neutron lifetime. It is in this way that the state of things is described on page PDG devoted to neutron lifetime [2]. The best way of solving this problem is to conduct new more precise experiments. One could also make a more detailed analysis of earlier experiments and seek for possible systematic errors.

---

* Corresponding author. Tel.: +7-81371-46001; fax: +7-81371-30072.
*E-mail address*: serebrov@pnpi.spb.ru.

Fig. 1 and Table 1 show dynamics of developing events. Before making measurements [1] by "Gravitrap" installation the world time was mainly determined by the results of paper [5]. At that time there was obtained the non discrepant world average value 885.7 ± 0.8 s. A new precise measurement of neutron lifetime in 2004 [1] gave rise to the discrepancy described above. The discrepancy became more obvious in 2007 after obtaining the results of measurements of neutron lifetime by a UCN magnetic trap [3]. It is easy to see that the experiment [5] is one of the most precise experiments in Table 1. It makes the main contribution into the world average value obtained till 2004, it also makes the main contribution into the divergence of data in previous and new measurements.

First of all we have made the analysis of our experiment. We have created the Monte Carlo model and could not find any systematic errors in our measurements. The results of these analyses have been published in paper [1]. The systematic error 7 s in our measurements, in which extrapolation of UCN storage time to the neutron lifetime comprises only 5 s, seems to be unlikely. The next stage of our analysis is the analysis of the experiment [5], where extrapolation is 100-120 s and it is confirmed that it has been done with systematic uncertainty 0.4 s. It is this assumption that is quite doubtful. A detailed analysis of the experiment [5], conducted with the Monte Carlo model is made in our paper [21]. We have also made the analysis of the experiment MAMBO I [10] in our paper [22]. In both experiments we have found a negative correction to the neutron lifetime approximately equal to 6 s. The Monte Carlo simulation has been done with the set of programs aimed at simulation of UCN experiments [23]. Verification of the set of programs has been fulfilled by comparing the calculation of results with that made by GEANT4 modified for UCN [24]. In addition, testing has been made on experimental data in our papers [1] and [22].

The storage vessel in the experiment [5] consists of two coaxial cylinders. The storage of UCN is possible either in the internal cylinder or in the gap between two cylinders. It results either in altering the relationship between the surface and the volume or in altering the relationship between the fraction of neutrons having decayed in the time of flight to that of the neutrons lost in the walls. Later on there is an extrapolation to the neutron lifetime using the rate of counting the neutrons heated inside the vessel and registered by detectors $^3$He, located around the storage chambers. The analysis made in our paper [21], revealed two most meaningful errors. One of them is caused by non-equivalence of UCN registration after storage either in internal or external vessel. In the case of storage in the internal vessel the valves open inwards of the volume and heat the fraction of neutrons located there. Such an effect does not occur in the external vessel. This effect is observed in rough experimental data as a peak in the diagram of detector counting the heated neutrons for an internal vessel, and is not observed for an external vessel [25]. This effect is not considered in papers [5] and [25]. Our simulation enabled us to make the evaluation of the effect. Insufficient knowledge of the character of the valve movement gives rise to uncertainty as far as the value of this effect is concerned. The second systematic error arises because of unequal efficiency of detector for heated neutrons in storing in external and internal vessels. As in the previous case, here is uncertainty of correction concerned with the uncertainty of effective length of neutron registration by a detector of thermal neutrons. For the claimed precision to be obtained, one needs a detailed calibration of detectors. Unfortunately in paper [5] this effect had the wrong sign, which requires an additional correction. Due to the above mentioned uncertainties the claimed systematic experimental error [5] 0.4 s appears to be problematic even after taking account of the registered effects. A detailed analysis of systematic errors is given in paper [21]. After accepting a correction and an error from this paper, the result of neutron lifetime from [5] will be 879.9 ± 0.9$_{stat.}$ ± 2.4$_{syst}$ s. The corrected result is in agreement with the result 878.5 ± 0.8 s from [1].

Table 1. Progress of neutron lifetime measurements till 2007.

| $\tau_n$, s | Author(s), year, reference |
|---|---|
| 878.2 ± 1.9 | V. Ezhov et al. 2007 [3] |
| 878.5 ± 0.7 ± 0.3 | A. Serebrov et al. 2005 [1] |
| 886.3 ± 1.2 ± 3.2 | M.S. Dewey et al. 2003 [4] |
| 885.4 ± 0.9 ± 0.4 | S. Arzumanov et al. 2000 [5] |
| 889.2 ± 3.0 ± 3.8 | J. Byrne et al. 1996 [6] |
| 882.6 ± 2.7 | W. Mampe et al. 1993 [7] |
| 888.4 ± 3.1 ± 1.1 | V. Nesvizhevski et al. 1992 [8] |
| 893.6 ± 3.8 ± 3.7 | J. Byrne et al. 1990 [9] |
| 887.6 ± 3.0 | W. Mampe et al. 1989 [10] |
| 872 ± 8 | A. Kharitonov et al. 1989 [11] |
| 878 ± 27 ± 14 | R. Kossakowski et al. 1989 [12] |
| 877 ± 10 | W. Paul et al. 1989 [13] |
| 891 ± 9 | P. Spivac et al. 1988 [14] |
| 876 ± 10 ± 19 | J. Last et al. 1988 [15] |
| 870 ± 17 | M. Arnold et al. 1987 [16] |
| 903 ± 13 | Y.Y. Kosvintsev et al. 1986 [17] |
| 937 ± 18 | J. Byrne et al. 1980 [18] |
| 881 ± 8 | L. Bondarenko et al. 1978 [19] |
| 918 ± 14 | C.J. Christensen et al. 1972 [20] |

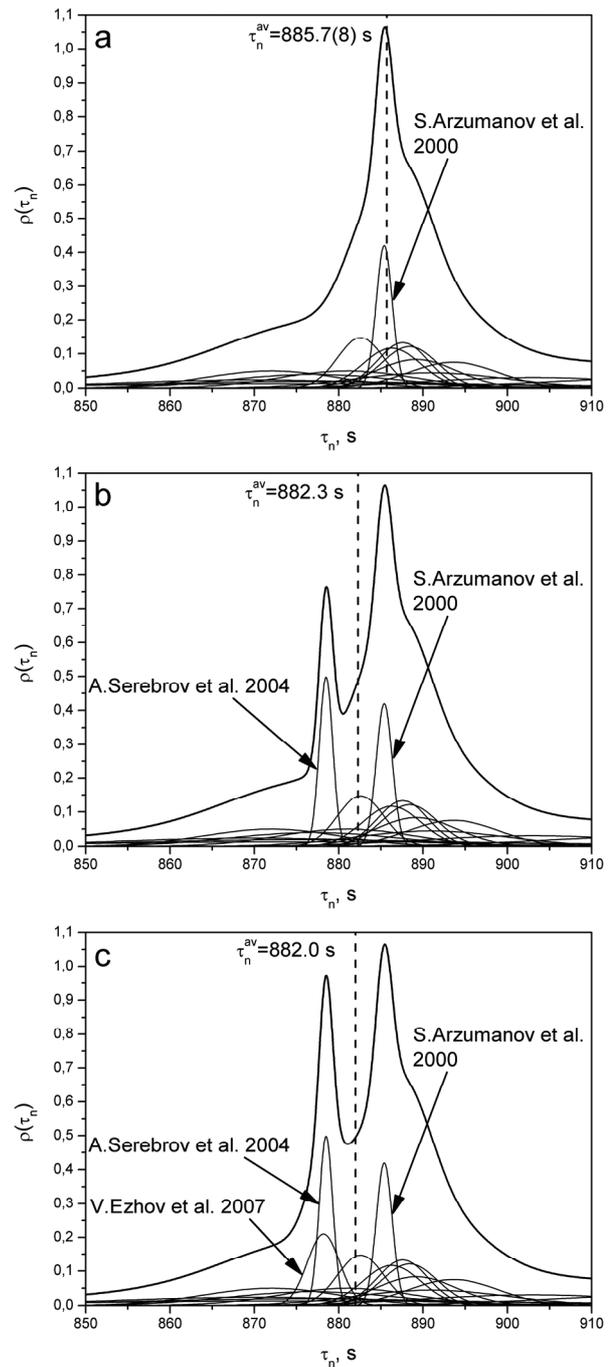

Figure 1. Progress of the neutron lifetime measurements. a: before "Gravitrap" measurement in 2003; b: after "Gravitrap" measurement in 2004; c: after magnetic trap measurement in 2007.

The next stage of our analysis was the experiment [10]. Since after completing the experiment [10] one discovered the effect of quasi-elastic scattering of UCN on the surface of the liquid fomblin, we created the Monte Carlo model taking into consideration this process. As a result two systematic effects were found. One of them is concerned with long storing of above-barrier neutrons while the other one is caused by quasi-elastic scattering of UCN on the surface of the liquid fomblin. A detailed analysis of these effects is available in our paper [22]. The discovered correction resulting from the effects of above-barrier neutrons and quasi-elastic scattering is –6.0 ± 1.6 s. Systematic experimental error [10] is about 3 s. It considerably makes up for lack of information on experimental details. The corrected result is in agreement with the result 878.5 ± 0.8 s [1].

Now one should give a new table of data on neutron lifetime measurements with corrections from papers [5] and [21] as well as from papers [10] and [22]. We also added experimental result MAMBO II [26] to the table, which is continuation of the experiment MAMBO I. The MAMBO II used the spectrum without above-barrier neutrons. Therefore the systematic of the experiment MAMBO I was suppressed.

The paper [8] can be withdrawn from the list, since much more accurate data have been obtained by this installation using low-temperature fomblin rather than solid oxygen. The difference between earlier and new data is 2.9 standard deviations. We assume that covering made of solid oxygen is not so reliable and that the covering of a narrow trap is more problematic. The difference in 2.9 standard deviations can be caused by this circumstance.

Table 2. The table of the experimental results for the neutron lifetime after corrections and additions.

| $\tau_n$, s | Author(s), year, reference |
|---|---|
| 881.5 ± 2.5 | S. Arzumanov et al. 2009 [27][*] |
| 878.2 ± 1.9 | V. Ezhov et al. 2007 [3][*] |
| 878.5 ± 0.7 ± 0.3 | A. Serebrov et al. 2005 [1][*] |
| 886.3 ± 1.2 ± 3.2 | M.S. Dewey et al. 2003 [4] |
| 879.9 ± 0.9 ± 2.4 | S. Arzumanov et al. 2000 [5,21][*] |
| 880.7 ± 1.8 | A. Pichlmaier et al. 2010 [26][*] |
| 889.2 ± 3.0 ± 3.8 | J. Byrne et al. 1996 [6] |
| 882.6 ± 2.7 | W. Mampe et al. 1993 [7][*] |
| 893.6 ± 3.8 ± 3.7 | J. Byrne et al. 1990 [9] |
| 881.6 ± 3.0 | W. Mampe et al. 1989 [10,22][*] |
| 872 ± 8 | A. Kharitonov et al. 1989 [11][*] |
| 878 ± 27 ± 14 | R. Kossakowski et al. 1989 [12] |
| 877 ± 10 | W. Paul et al. 1989 [13][*] |
| 891 ± 9 | P. Spivac et al. 1988 [14] |
| 876 ± 10 ± 19 | J. Last et al. 1988 [15] |
| 870 ± 17 | M. Arnold et al. 1987 [16] |
| 903 ± 13 | Y.Y. Kosvintsev et al. 1986 [17][*] |
| 937 ± 18 | J. Byrne et al. 1980 [18] |
| 881 ± 8 | L. Bondarenko et al. 1978 [19] |
| 918 ± 14 | C.J. Christensen et al. 1972 [20] |

[*] UCN experiments

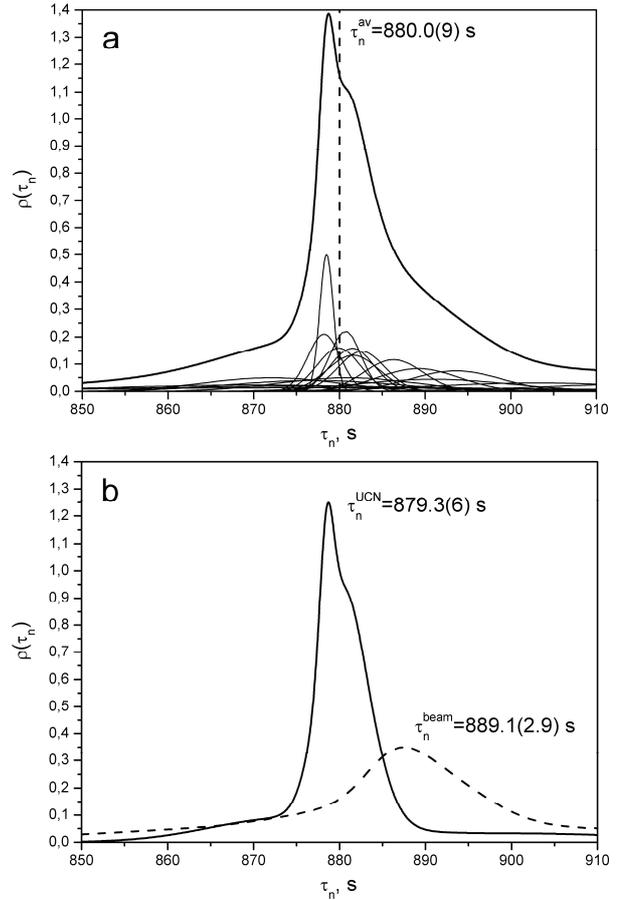

Figure 2. a) Distribution of results of measurements for the neutron lifetime after corrections and additions, giving average value of 880.0 ± 0.9 s. b) Distribution of two groups of data: experiments on neutron lifetime with UCN and beam experiments.

Finally the table should contain new data obtained by V.I. Morozov's group [27], published at the conference. Then after corrections and additions the table of experimental data on neutron lifetime looks in the following way (Table 2, Fig. 2). The standard error of the average value of neutron lifetime according to Table 2 is 0.6 s, while the standard deviation is 0.9 s. Thus it will be reasonable to accept 880.0 ± 0.9 s as the world average value for neutron lifetime.

To summarize, one should note that the analysis of neutron β-decay with a new world average value for neutron lifetime is in agreement with Standard Model. This analysis is shown in Fig.3 and described in detail in papers [28,29]. Fig. 3 shows dependence of matrix element of quark mixing $|V_{ud}|$ on axial coupling constant $g_A$ at different values of neutron lifetime. The value $|V_{ud}|$ = 0.9743(7), calculated for a new world average value of neutron lifetime 880.0(9) s and $g_A$ = 1.2750(9) [30], agrees with $|V_{ud}|$ = 0.97419(22), calculated from unitarity of the quark mixing matrix [2] and $|V_{ud}|$ = 0.97425(22), measured from nuclear β-decay [30,31]. One can see that the value $|V_{ud}|$ = 0.9711(6), obtained for an earlier world average value of neutron lifetime 885.7(8) s, does not coincide with experimental data $|V_{ud}|$ = 0.97419(22) and $|V_{ud}|$ = 0.97425(22).

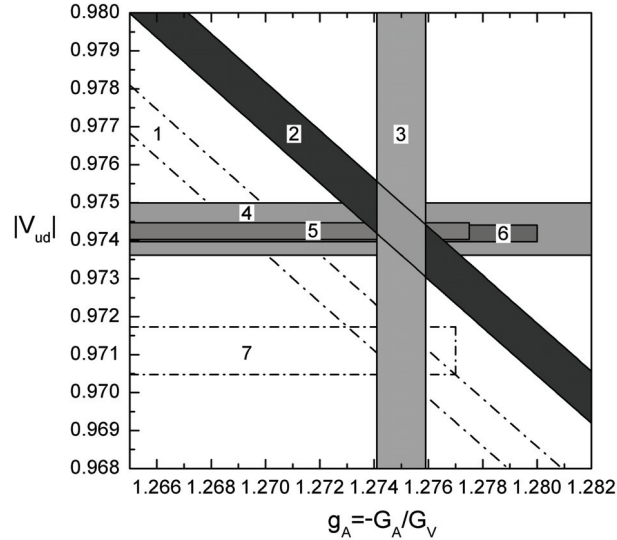

Figure 3. The dependence of the CKM matrix element $|V_{ud}|$ on the values of the neutron lifetime and the axial coupling constant $g_A$. (1) neutron lifetime, PDG 2006; (2) neutron lifetime, this article; (3) neutron β-asymmetry, Perkeo 2007; (4) neutron β-decay, this article + Perkeo 2007; (5) unitarity; (6) $0^+ \to 0^+$ nuclear transitions; (7) neutron β-decay, PDG 2006 + Perkeo 2007.

Besides it is to be noted, that a comprehensive analysis has been made of original nuclear synthesis at the early stage of the Universe formation [32]. There has been analyzed the influence of a new value of neutron lifetime on the agreement of data on original extension of D and $^4$He with the data on barion asymmetry $\eta_{10}$. Application of a new value of neutron lifetime improves the data agreement on original extension of D, $^4$He and barion asymmetry. Although the accuracy of cosmological data is much lower than that of measurements of neutron lifetime, the shift between the earlier average world value and a new one undoubtedly affects the verification of nuclear synthesis model at the original stage of the Universe formation.

Before completing the paper we would like to make a separate analysis of experiments on neutron lifetime with UCN and beam experiments. The average neutron lifetime with UCN is equal to 879.3(0.6) s, while for beam experiments it is 889.1(2.9) s. Fig. 2b illustrates the distribution of two groups of data. Fig. 4 shows two groups of data in the historical order.

The available difference between average values of two groups is (3.3σ) and needs consideration. The contribution of beam experiments into the world average value is not high, therefore it does not affect the analysis made above. However it is an independent problem that must be solved. It is highly desirable that the precision of beam experiments should be enhanced.

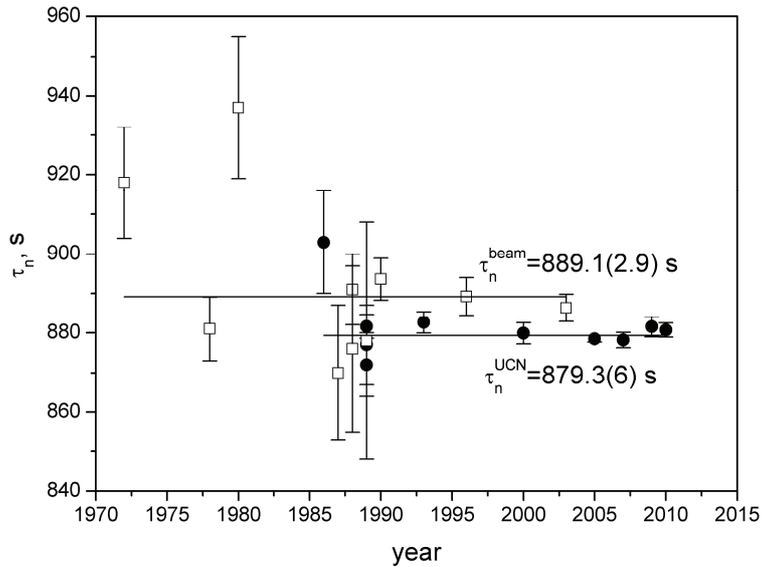

Figure 4. Two groups of data in the historical order: ● – experiments on neutron lifetime with UCN, □ – beam experiments.

This research has been carried out at the support of the Russian Fund of fundamental investigations, grants № 07-02-00859-a, 08-02-01052-a, 10-02-00217-a, 10-02-00224-a. It has also been supported by the Ministry of education and science of Russian Federation. State contracts № P2427, P2500, P2540, 02.740.11.0532, 14.740.11.0083.